# A HIGGS-FREE MODEL FOR FUNDAMENTAL INTERACTIONS⋆ PART II: Predictions for Electroweak Observables

**Marek Pawłowski**[1†]
Soltan Institute for Nuclear Studies, Warsaw, POLAND
**and**
**Ryszard Rączka**[2‡]
Soltan Institute for Nuclear Studies, Warsaw, POLAND
and
Interdisciplinary Laboratory for Natural and Humanistic Sciences
International School for Advanced Studies (SISSA), Trieste, ITALY

May, 1995

**Abstract**

The predictions for electroweak observables following from Higgs-Free Model for Fundamental Interactions are derived. It is shown that these predictions are close to the Standard Model predictions and they are in a surprising agreement with the experimental data. The analysis of electroweak observables at low and high energy suggests that the Higgs mass $m_H$ is just the ultraviolet cutoff which increases if the process energy scale increases. We propose several experimental tests which can discriminate between the Standard and the Higgs-Free Models.

⋆ This paper is the refined version of the preprint ILAS/EP-1-1995 (hep-ph 9501370), which takes into account the newest electroweak data and is now published in two parts: Part I: Formulation of the model, Part II: Predictions for electroweak observables.

[1] Partially supported by Grant No. 2 P302 189 07 of Polish Committee for Scientific Researches.
[2] Partially supported by the Stiftung Für Deutsch-Polnische Zusammenarbeit Grant No. 984/94/LN.

† e–mail: PAWLOWSK@fuw.edu.pl
‡ e–mail: RRACZKA@fuw.edu.pl

# 1 Introduction

We have proposed in Part I of our work (referred here as I) the Higgs-Free Model (HFM) for fundamental interactions [1]. This model – in the limit of flat space-time – represents a massive vector boson model for electroweak and strong interactions, which is perturbatively nonrenormalizable [2],[3],[4]. Hence there is a problem of getting predictions for observables from such a model.

The conventional Standard Model (SM) is tested presently up to one-loop radiative corrections [5]. Hence it is natural that we shall derive predictions from HFM up to one-loop and compare them with the relevant predictions of SM. Fortunately the method for a construction of the one-loop effective lagrangian for various nonrenormalizable models with the same gauge symmetry as the SM symmetry was developed recently [6],[7]. Hence one has in principle a well elaborated formalism for a derivation of predictions for observables in HFM in one-loop approximation.

We have shown in Sec. 6 of I that if we choose the ultraviolet (UV) cutoff $\Lambda$ in HFM equal to the mass $m_H$ of Higgs particle then HFM and SM predictions almost coincide and they become equal in the limit $m_H \to \infty$. This observation is supported also by the experimental fact that if the energy scale of considered electroweak processes increases then the central value of the Higgs mass deduced from these processes also increases [1]. Thus it looks like that the Higgs mass $m_H$ plays in fact in SM the role of UV cutoff. We present now in Sec. 2 a method for for a determination of predictions for observables in HFM. We show in the Appendix that this method is equivalent to the method of one loop effective lagrangian presented in [6] if one chooses in the last approach a definite renormalization scheme.

We give in Sec. 3 the HFM predictions for electroweak lepton observables which are measured in LEP and SLD experiments. Next we present in Sec. 4 our predictions for electroweak hadron and mixed hadron-lepton observables which are measured in LEP experiments. It is noteworthy that in spite of the absence of the Higgs particle in HFM our predictions for electroweak lepton and hadron observables agree with experimental data very well and they are close to SM predictions. Hence it seems that one does not need a Higgs particle for explaining the existing experimental data.

We discuss in Sec. 5 a possibilities of experimental discrimination between SM and HFM. We complete our work with a discussion of obtained results and an enumeration of open theoretical and experimental problems connected with various aspects of HFM and the conventional SM.

# 2 Derivation of predictions for electroweak observables.

Our model represents in fact the gauge field theory model with massive vector mesons and fermions. It is well–known that such models are in general nonrenormalizable [2]-[4]. We remind however that in the nonrenormalizable Fermi model for weak interactions we can make a definite predictions for low energy phenomena e.g. for $\mu$ or neutron decays. The effective method of getting the definite predictions for observables from the nonrenormalizable massive vector meson gauge field theories was recently developed by Herrero and Morales [6] and by Bilenky and Santamaria [7]. This method consist on a derivation of the effective lagrangian corresponding to the L-loop approximation of the original nonrenormalizable model from which one can derive the finite expression for any observable of the considered theory [7],[8]. In that manner Herrero and Morales derived one-loop expressions for the observables $\Delta r$ and $\Delta\rho$ for a simplified electroweak model in which fermions are disregarded [6]. Similarly the recent progress with the so called Generalized Equivalence Theorem allows to make definite predictions for the probability amplitudes in nonrenormalizable models like gauged nonlinear $\sigma$–model or other nonrenormalizable gauge field theory models [9]. Hence in our HFM we can obtain definite predictions for electroweak phenomena if we consider processes with energy scale $E$ below some ultraviolet (UV) cutoff $\Lambda$. We have shown in Section 6 of I that the cutoff $\Lambda$ is closely connected with the Higgs mass $m_H$ appearing in the Standard Model. Hence, from this point of view, Higgs mass is nothing else as the UV cutoff which assures that the truncated perturbation series is meaningful.

We would like to discuss now the problem of getting predictions from our nonrenormalizable model for quantities like $W$-meson mass $m_W$, $sin^2\theta_W^{eff}$ of effective Weinberg angle, lepton width $\Gamma_l$ and other characteristics of $Z$ peak in $e^+e^-$ collision which are measured in so called precision tests of electroweak theories. It is known that the SM predictions for these quantities including one-loop radiative corrections depend on the unknown value of the Higgs mass. On the contrary, the one loop-predictions of our model depend on the cutoff parameter $\Lambda$. One can calculate these predictions directly or one can use SM results and correct them using Equation (6.3) of I [10].

In order to obtain the definite predictions we have to select some "EW-meter" i.e. a quantity $R_0(\Lambda)$ which is measured with the best accuracy in the present EW experiments and which will replace an unknown variable $\Lambda$ in the expressions for the other physical quantities $R_i$. Inverting the relation $R_0(\Lambda)$ we can express $\Lambda$ as the function of $R_0$: $\Lambda = \Lambda(R_0)$. Then we insert this relation into the expression for all other observables $R_i$ and we get the cutoff-independent definite function

$$R_i = \tilde{R}_i(R_0) = R_i(\Lambda(R_0)) \tag{2.3}$$

as the prediction of our model. We show in the Appendix that the proposed method of derivation of predictions for observables determines a choice of definite renormalization scheme for the effective field theory connected with our nonrenormalizable model.

We assume that the best candidate for "EW-meter" must fulfill the following criteria:

i) should be measured directly (what excludes $sin^2\theta_W^{eff}$ obtained combining results of different asymmetries),

ii) should be measured with best accuracy relatively to the slope of its $\Lambda$ dependence what means that the ratio

$$\frac{\text{experimental error of } R_0}{dR_0/d\Lambda} \tag{2.4}$$

must be minimal at the measured central value of $R$.

The numerical analysis indicates that the "EW-meter" is presently given by the total width of $Z$-meson $\Gamma_Z = 2497.4 \pm 3.8$.

We calculated within our model the $\Gamma_Z$-dependence (2.3) for almost all quantities measured in the precision tests of EW theories and we deduced from these functions HFM predictions for these observables.

We shall discuss now separately the leptonic and the hadronic observables.

# 3 HFM predictions for leptonic observables.

The leptonic observables are better determined by the theory since in general they are almost independent from the strong coupling constant $\alpha_s$ and one does not encounter in the process of calculation the problem of a hadronization of QCD partons into final hadrons. In addition, as we will see below, these predictions in our model are only weakly dependent from the top quark mass $m_t$.

We give in Fig. 3.1 the plot of six lepton electroweak observables as the function of $\Gamma_Z$ obtained in our model after elimination of the ultraviolet cutoff $\Lambda$. The quantities are:

– $sin^2\theta_W^{eff}$ – the sinus squared of the effective Weinberg angle,

– $m_W/m_Z$ – the ratio of weak boson masses,

– $\Gamma_l$ – the partial $Z_0$ width of the decay into one lepton family,

– $\mathcal{A}_l^0 = \frac{\bar{g}_{L_l}^2 - \bar{g}_{R_l}^2}{\bar{g}_{L_l}^2 + \bar{g}_{R_l}^2} = \frac{2\bar{g}_{V_l}\bar{g}_{A_l}}{\bar{g}_{V_l}^2 + \bar{g}_{A_l}^2}$ – the asymmetry of left and right vector-meson-lepton couplings,

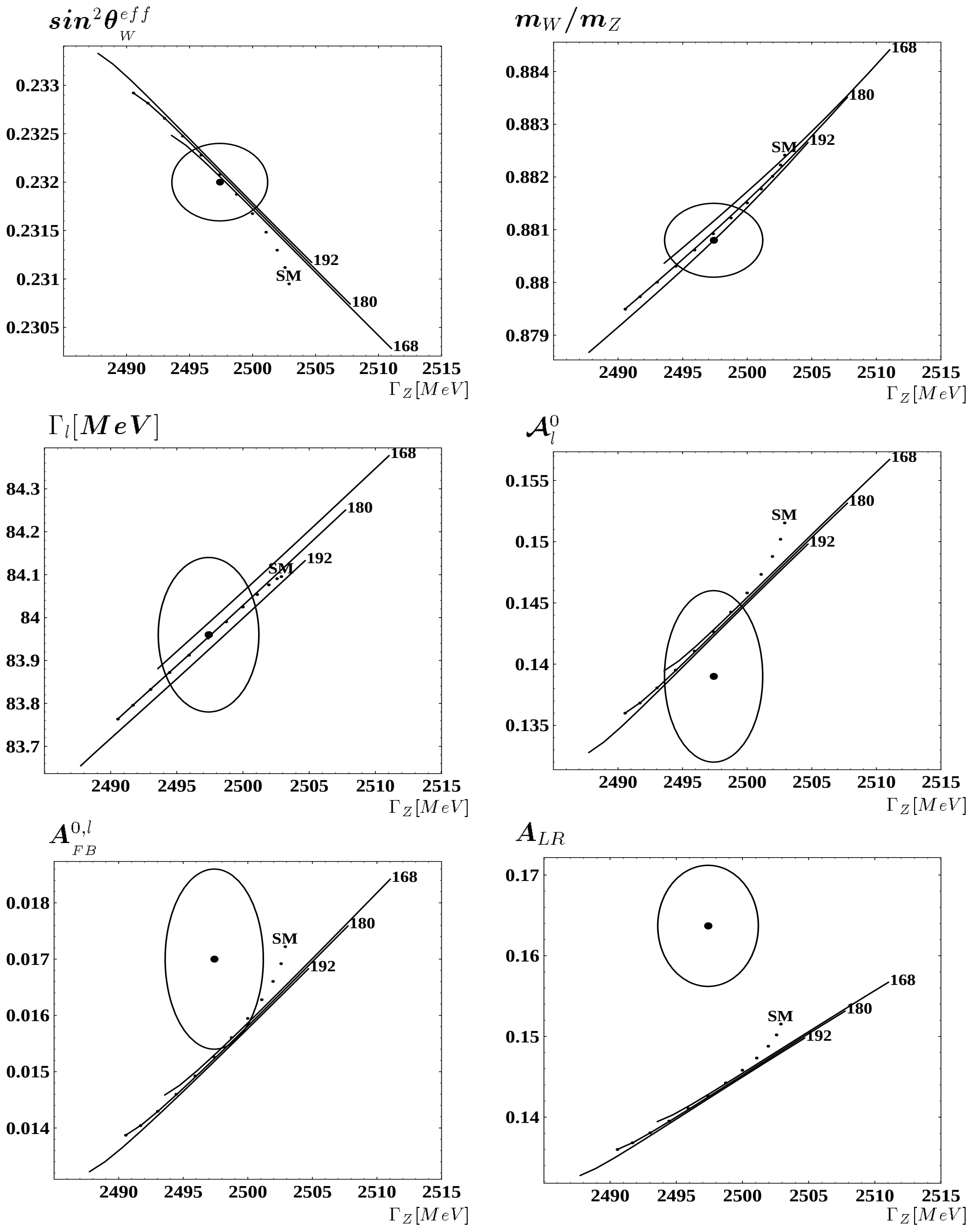

**Fig. 3.1** *The plots of six lepton electroweak observables as the functions of our "EW-meter" $\Gamma_Z$. We give for each quantity tree curves corresponding to top masses $m_t = 168GeV$, $m_t = 180GeV$ and $m_t = 192GeV$ respectively and a dotted curve which represents SM prediction for $m_t = 180GeV$ and $m_H$ varying logarithmically from 50GeV (right side of the curve) to 2.25TeV. The circles are plotted at the central experimental value and the ellipses are plotted at 1 standard deviation.*

– $A_{FB}^{0,l}$ – the forward-backward asymmetry of lepton pair production at $Z_0$ peak,

– $A_{LR}$ – the left-right asymmetry of $Z_0$ production from longitudinal polarized beam.

The three continuous curves at each plot correspond to our predictions as the function of $\Gamma_Z$ for various top masses. The ball represents the central experimental values for considered observable and $\Gamma_Z$ and the ellipse is plotted at one standard deviation from these values. The dotted curves represent the SM predictions for $m_t = 180 GeV$. We use in all calculations the set of experimental data from all LEP groups available in February 1995 [11]; in particular we take as the input $\alpha_s = 0.125$, $\alpha^{-1}(m_Z) = 128.87$ and $m_Z = 91.1888 GeV$. We use in calculations of all observables the correlation matrix from [11]. We used for computation the ZFITTER programme v4.8 (of 07.09.94).

We see that $\Lambda$ cutoff independent predictions from our model almost coincide with SM predictions and for most of quantities agree surprisingly well with experimental data. Our predictions for lepton observables are almost independent from the present uncertainty as to the value of top mass. In Table 3.1 we have collected the experimental values, HFM predictions and SM predictions for the listed six lepton electroweak observables.

| Observable | Experimental value | HFM prediction | SM prediction |
|---|---|---|---|
| $sin^2\theta_W^{eff}$ | $0.2320 \pm 0.0003^{+0.0000}_{-0.0002}$ | $0.2321 \pm 0.0006$ | $0.2322 \pm 0.0003 \pm 0.0004$ |
| $m_W/m_Z$ | $0.8808 \pm 0.0007$ | $0.8808 \pm 0.001$ | $0.8807 \pm 0.0002 \pm 0.0007$ |
| $\Gamma_l[MeV]$ | $83.96 \pm 0.18$ | $83.94 \pm 0.08$ | $83.90 \pm 0.02 \pm 0.19$ |
| $\mathcal{A}_l^0$ | $0.139 \pm 0.007$ | $0.142 \pm 0.004$ | $0.142 \pm 0.003 \pm 0.003$ |
| $A_{FB}^{0,l}$ | $0.0170 \pm 0.0016$ | $0.0151 \pm 0.0008$ | $0.0151 \pm 0.0005 \pm 0.0006$ |
| $A_{LR}$ | $0.1637 \pm 0.0075$ | $0.142 \pm 0.004$ | $0.142 \pm 0.003 \pm 0.003$ |

**Tab. 3.1** *Comparison of experimental values, HFM predictions and SM predictions for lepton electroweak observables.*

# 4 HFM predictions for hadronic observables.

Most of hadronic observables are strongly dependent on the strong coupling constant $\alpha_s$ and the top quark mass $m_t$. We give in Fig. 4.1 the plot of six hadronic electroweak observables as the function of $\Gamma_Z$ obtained in our model after elimination of the ultraviolet cutoff $\Lambda$. The quantities are:

– $\sigma_h^0 = \frac{12\pi}{m_Z^2}\frac{\Gamma_e\Gamma_{had}}{\Gamma_Z^2}$ – the hadronic pole cross section,

– $R_l = \frac{\Gamma_{had}}{\Gamma_l}$,

– $R_b = \frac{\Gamma_b}{\Gamma_{had}}$,

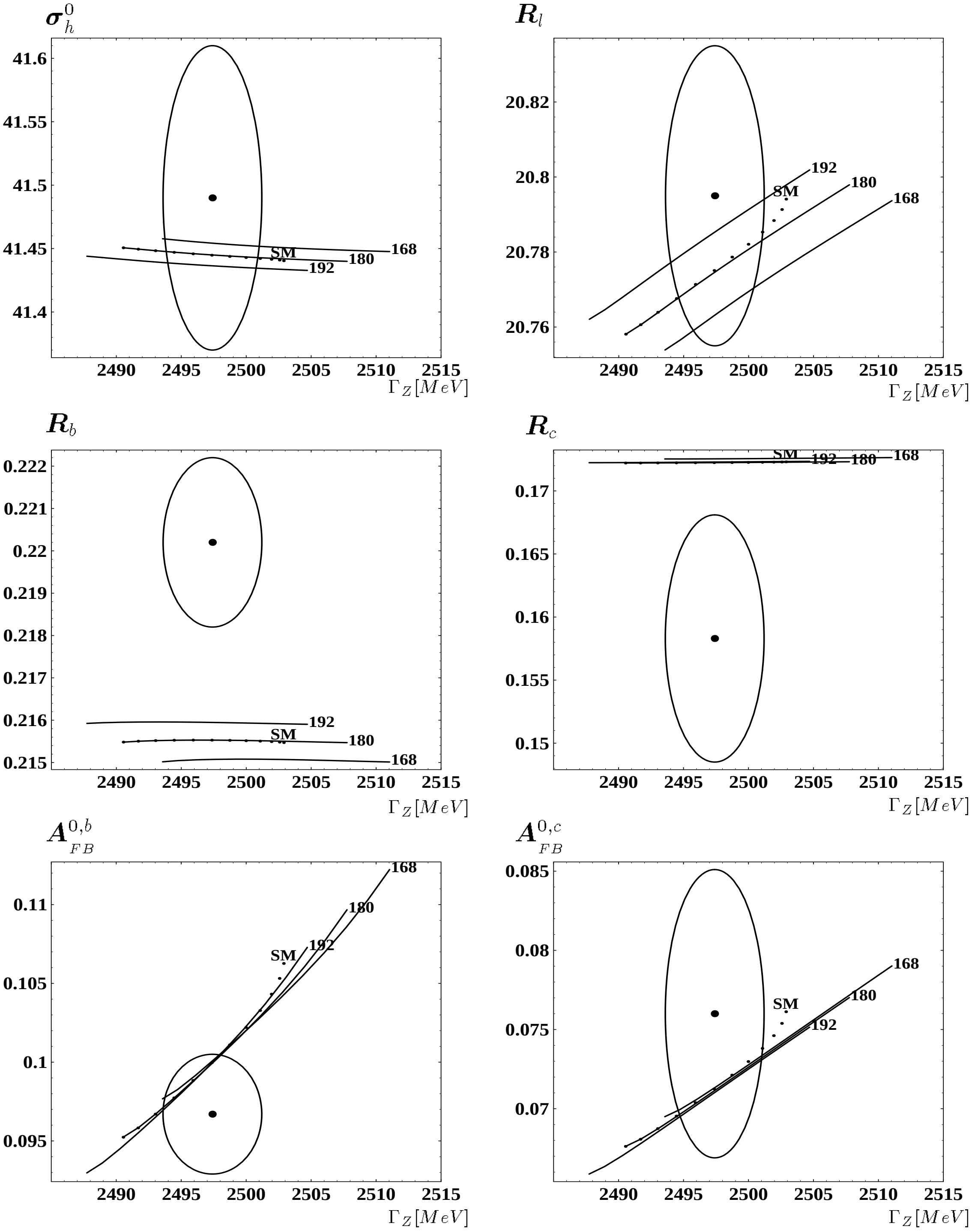

**Fig. 4.1** *The plots of six hadron electroweak observables as the functions of our "EW-meter" $\Gamma_Z$.*

– $R_c = \frac{\Gamma_c}{\Gamma_{had}}$
– $A_{FB}^{0,b}$ – the forward-backward asymmetry of $b\bar{b}$ pair production at $Z_0$ peak,
– $A_{FB}^{0,c}$ – the forward-backward asymmetry of $c\bar{c}$ pair production at $Z_0$ peak.

We present in Table 4.1 the comparison of experimental results for hadronic observables with HFM and SM predictions.

| Observ. | Exp. value | HFM prediction | SM prediction |
|---|---|---|---|
| $\sigma_h^0[nb]$ | $41.49 \pm 0.12$ | $41.44 \pm 0.01 \pm [0.02]$ | $41.44 \pm 0.004 \pm 0.01 \pm [0.02]$ |
| $R_l$ | $20.795 \pm 0.040$ | $20.78 \pm 0.009 \pm 0.016 \pm [0.03]$ | $20.784 \pm 0.006 \pm 0.003 \pm [0.03]$ |
| $R_b$ | $0.2202 \pm 0.0020$ | $0.2156 \pm 0.0004$ | $0.2156 \pm 0 \pm 0.0004$ |
| $R_c$ | $0.1583 \pm 0.0098$ | $0.171 \pm 0$ | $0.171 \pm 0 \pm 0$ |
| $A_{FB}^{0,b}$ | $0.0967 \pm 0.0038$ | $0.0995 \pm 0.0025$ | $0.0994 \pm 0.002 \pm 0.002$ |
| $A_{FB}^{0,c}$ | $0.0760 \pm 0.0091$ | $0.071 \pm 0.002$ | $0.071 \pm 0.001 \pm 0.001$ |

**Tab. 4.1** *Comparison of experimental values, HFM predictions and SM predictions for hadron electroweak observables.*

We see a remarkable agreement of HFM predictions with experimental data.

# 5 Discussion.

It follows from Tables 3.1 and 4.1, and from equation (6.2) of I that HFM gives predictions for EW observables which are very close to SM predictions and which surprisingly well describe the experimental data. We are in the process of deriving the predictions for other electroweak data – for instance in the low or intermediate energy regions – but it is almost evident that also in these cases the HFM will be equally successful. There arises therefore an interesting problem how to discriminate between HFM and SM.

We see two possibilities:

I. Consider the EW processes in which the SM Higgs particle gives the resonant contribution. The analysis of such processes like Z-meson pair or $W^+W^-$ production in gluon-gluon or photon-photon scattering is planned in LEP200, LHC or NLC experiments [12, 13, 14]. In these cases the SM gives the characteristic pick in the cross section considered as the function of invariant mass of the produced particles [12, 13, 14]; contrary the HFM gives the pickless mononotonic cross section decreasing with increasing invariant mass. The expected position of SM Higgs resonant pick can be estimated from the informations provided by EW precision tests. At present these informations are still not very conclusive and essentially depend on the assumptions concerning the set of fitted experimental data [15]. We can expect, however that this situation will soon improve mostly due to the increasing accuracy of the top and W mass measurements and the further analysis of LEP1 and SLAC data. The constructed LEP2 and LHC accelerators will be probably able to test directly the region of the Higgs mass predicted by the precision tests. The possible negative result of the direct searches for the Higgs particle with the mass $m_H$ in the interval admissible by EW precision tests would represent an evident contradiction in the frame-work of SM. The above contradictions will not appear in HFM since there the Higgs particle is absent.

II. Consider the electroweak processes at various energy scales. We have shown in I. Sec. 6 that all the present low energy data give the $sin^2\theta_W$ of the Weinberg angle smaller than the high energy experiments [5, 11]. This implies that the admissible Higgs mass interval read off from the high energy experiments is smaller than the corresponding interval from the low energy experiments. If the planned increase of the precission of low energy experiments [16] will imply that the admissible Higgs mass intervals from the low and the high energy experiments will be disjoint we obtain a contradiction within the frame-work of SM. This contradiction will not appear in HFM since the Higgs particle is absent in this model and the Higgs mass parameter is replaced by the UV cutoff $\Lambda$ which may change with the energy scale.

We see therefore that expected in a near future considerable increase of accuracy of EW precision tests in high and low energy region and the new accelerator experiments may provide a crucial tests for Standard and Higgs-Free Models.

ACKNOWLEDGMENTS

The authors are grateful to G. Altarelli, I Białynicki Birula, B. Grządkowski, Z. Haba, M. Kalinowski, J. Werle and G. Veneziano for interesting discussions and S.D. Odintsov for sending to us the results of his group. They are especially grateful to S. Dittmaier for sending them his computer code and to D. Schildknecht for the extensive discussion of properties of his model.

# Appendix.

We have proposed in Section 2 a method of getting of the cutoff independent predictions from our model. We show now that this method is equivalent to a selection of a definite renormalization scheme in our model. For the sake of simplicity we shall present our arguments in case of nonrenormalizable $\sigma$-model obtained from the SM in the limit $m_H \to \infty$ restricted to the boson sector only. This model was considered in detail by Herrero and Morales (H-M) in the series of papers [6]. We shall consider for an illustration two basic electroweak radiative corrections $\Delta r$ and $\Delta\rho$. In our method, using the dimensional regularization we have the following cutoff dependent representation for these quantities:

$$\Delta r = \Delta r_F + k_r \Delta_\varepsilon; \quad \Delta\rho = \Delta\rho_F + k_\rho \Delta_\varepsilon.$$

Here $\Delta r$ and $\Delta\rho$ represents the finite cutoff independent part of $\Delta r$ and $\Delta\rho$ respectively in which all conventional one-loop renormalizations of coupling constants masses etc... were carried out and

$$\Delta_\varepsilon = \frac{2}{\varepsilon} - \gamma_E + \ln 4\pi, \quad \varepsilon = 4 - D. \qquad (A.1)$$

The term $k\Delta_\varepsilon$ is the UV divergent term which in the conventional SM is canceled by the contribution from Feynman diagrams with the Higgs internal lines (absent in our and H-M models). Using the method proposed by us in Sec. 2 we get the cutoff independent relation between observables

$$\Delta\rho = \frac{k_\rho}{k_r}\Delta r + \Delta\rho_F - \frac{k_\rho}{k_r}\Delta r_F. \qquad (A.2)$$

We now show that using the full analysis of H-M we get the same relation if we choose a definite renormalization scheme (RS) which we shall specify.

We recall that the method of H-M is based on the technique of effective one-loop lagrangians associated with the given nonrenormalizable model. In H-M model the effective lagrangian is given by the formula

$$L_{H-M} = L_{NL} + \sum_{l=0}^{13} a_l^b L_l \qquad (A.3)$$

where $L_{NL}$ is the lagrangian of the original nonrenormalizable gauged nonlinear $\sigma$-model with $SU)2)_L \times U(1)$ symmetry and the second term contains fourteen counterterms of dimension four which assure that the calculated in one-loop approximation observables will be finite. The coefficients $a_l^b$ represent the bare unrenormalized coupling constants associated with counterterms [6]. H-M obtain the following formula for $\Delta r$ and $\Delta\rho$:

$$\Delta r = \Delta r_F + \sum_{l=0}^{13} r_l a_l + C_r,$$

$$\Delta\rho = \Delta\rho_F + \sum_{l=0}^{13} \rho_l a_l + C_\rho. \qquad (A.4)$$

Here $r_l$ and $\rho_l$, $l = 0, ...13$, $C_r$, and $C_\rho$ are known, numerical coefficients and $a_l$, $l = 0, ..., 13$ are the renormalized coupling constants associated with counterterms. From (A.4) we obtain

$$\Delta\rho = \frac{k_\rho}{k_r}\Delta r + \Delta\rho_F - \frac{k_\rho}{k_r}\Delta r_F + \sum_{l=0}^{13}(\rho_l - \frac{k_\rho}{k_r} r_l)a_l + C. \qquad (A.5)$$

The quantities $a_l$ are depending on RS; one can see explicit example in the work by Gasser and Leutwyler [17],[6]. Therefore they cannot be calculated from the parameters like coupling constants,masses etc. of the original lagrangian $L_{NL}$ in (A.3). They can be determined only by the comparison of fourteen independent observables with the prediction of the form (A.4) of these observables. Comparing our expression (A.2) with the general H-M expression (A.5) we see that we get our result if we use as the one of the fourteen conditions determining the RS the condition

$$\sum_{l=0}^{13}(\rho_l - \frac{k_\rho}{k_r} r_l)a_l + C = 0. \qquad (A.6)$$

Expressing 14 independent observables $R_1$, ... $R_{14}$ in terms of say $R_0$ (equal e.g. to $\Delta r$) we shall get 14 conditions

$$\sum_{l=0}^{13}(\rho_l^n - \frac{k_\rho^n}{k_r^n} r_l^n)a_l = C_n, \quad n = 1, ...14. \qquad (A.7)$$

From (A.7) the coupling constants will be calculated and they will determine the RS chosen by our method.

Comparing our method with the conventional $\overline{MS}$ RS we see that we just cancel the divergent terms in observables; we see therefore that our method resemblance the $\overline{MS}$ RS applied on the level of observables.

If we would joint the fermion sector we would obtain in (A.3) more counterterms and the resulting new coupling constants $a_i^F$ coming from the fermion sector. However it is evident that our method will work in the same manner.